\documentclass[letterpaper,twocolumn]{esapub} 
          [1999/12/02 1.01 (PWD)]
\usepackage{times} 
\usepackage{graphicx}
\usepackage{natbib}

\title{LOWL p-mode frequencies and their variation with solar activity}
\author{Sebasti\'an J. Jim\'enez-Reyes$^{1,2}$}
\author{Thierry Corbard$^{1}$}
\author{Pere L. Pall\'e$^{2}$}
\author{Steve Tomczyk$^{1}$}
\affil{$^{1}$High Altitude Observatory, NCAR, PO Box 3000, Boulder, CO 80307 USA}
\affil{$^{2}$Instituto de Astrof\'\i sica de Canarias, E-38701, La Laguna,
Tenerife, Spain}

\begin{document}


\maketitle

\begin{abstract}
We present an analysis of the frequency shift and the even terms of the
frequency splitting coefficients carried out using six years of LOWL data, starting in 1994.
The temporal variations, and their dependences with the frequency
and degree are addressed. The results are consistent with previous
analysis.  
\end{abstract}

\section{Introduction}
It is very well established that the variations of some of the oscillation mode parameters is a  signature of the
solar activity cycle. For instance, the acoustic mode frequency is shifted
positively of 0.4$\mu$Hz from minimum to maximum of the solar cycle
(Libbrecht and Woodard
1990, R\'egulo  et al. 1994, Jim\'enez-Reyes  et al. 1998). 
It has also been noted that, this variation depends on the frequency and 
on the 
degree of the mode as well. The even terms of the splitting coefficients 
present similar changes. Here, we present an analysis of the frequency
shift using LOWL observations. 

\section{Data Analysis and Results}
The raw data consist in full disk solar Doppler images, which have been
collected by LOWL instrument. This experiment is based on a Magneto-Optical
Filter and has demonstrated its high sensitivity to the solar oscillations. 
The data were analyzed using the LOWL pipeline which
has  recently been improved (Jim\'enez-Reyes 2000). 
The images are first calibrated and then a spherical harmonic decomposition
 is
performed in order to create time series 
for degrees from $\ell$=0 to 99. Finally, a Fast
Fourier Transform is applied to the time series.

The spherical harmonics are not orthogonal over the
observed area, limited to one hemisphere. Therefore, the modes cannot be
totally isolated and some correlations between different ($\ell$,
$m$) spectra exist. Historically, the statistic of the power spectra has been
assumed to be as $\chi^2$ with 2 degrees of freedom. However, this
assumption is not correct when the spectra are not independent. 
Recently, important improvements have been achieved in the fitting techniques
to be applied to observations made using  instruments with spatial resolution (Schou 1992, Appourchaux et al. 1998). The real and the imaginary part
of the Fourier Transform follow a multi-normal distribution which is
described by a covariance matrix. The likelihood function
under these assumptions can be written as:
\begin{equation}
\label{likelihood_am}
S(\vec{a})= \sum_{i=1}^{N}{\log \mid E(\vec{a},\nu_{i})\mid +y^{T}(\nu_{i}) E(\vec{a},\nu_{i}) y(\nu_{i})} 
\end{equation}
where $y(\nu_{i})$ corresponds to the Fourier Transform and
$E(\vec{a},\nu_{i})$ is the covariance matrix  calculated by:
\begin{equation}
\label{emn}
E_{nm}(\vec{a},\nu_{i})= \sum_{m'=-\ell}^{\ell}{C_{nm'} C_{mm'}
v_{m'}(\vec{a},\nu_{i})} + B_{nm}  
\end{equation}
where $C$ is the leakage matrix which represents the correlation between
modes with different $m$ ($m$-leakage), whereas $B$ is the noise covariance 
matrix which gives information about the noise correlation between 
spectra. $v(\vec{a},\nu)$ is the variance and
it is given by a simple Lorentzian profile defined by the parameters $\vec{a}$
that we are trying to infer using a maximum likelihood method. The shift between $m$-components is  given by:
\begin{equation}
\label{Clebsch}
\nu_{n\ell m} = \nu_{n\ell} + \sum_{i=1}^{n_{coef}}{a_{i}(n,\ell) P_{i}^{\ell}(m)} 
\end{equation}
where $P_{i}^{\ell}(m)$ are orthogonal polynomials normalized such that
$P_{i}^{\ell}(\ell)=\ell$ (Schou et al. 1994, App. A). The odd $a$-coefficients
are induced by  the internal rotation while  the
even terms are mainly related to effects
of second order due to rotation, the presence of magnetic fields or any departure from spherical symmetry.

The distance between modes with the same order $n$
and with $\Delta\ell$=$\pm 1$ get closer and closer at high degrees leading to a leakage of the modes ($\ell$-leakage).
This problem arises at lower
degree in the case of observations from just one station, due to the existence
of the sidelobes as a result of the modulation of one day in the signal.
That is particular important for those modes where $\nu/L \approx 31 \mu$Hz. 
In order to reduce
the systematic errors an extra covariance matrix is added to take into
account the presence of spurious modes which can be close to the target
mode. Nevertheless, we have seen that the systematic errors are specially
important in the odd $a$-coefficients and not in the even ones.

At high degree, the numerical evaluation of the likelihood function get slow,
mainly because of  the large dimensions of the covariance matrix.
Therefore, we use just the main diagonal of the covariance
matrix. We have checked that this is indeed a very good approximation 
for high degrees.

\section{Frequency shift}

\begin{figure}[htbp]
\centering
\includegraphics[width=.7\linewidth,angle=90]{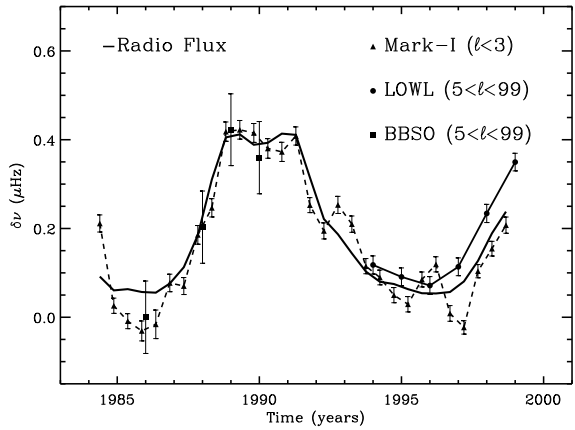}
\caption{Time variation of the integrated frequency shift. The BBSO
data have been averaged in the same way and plotted in the figure.
In addition, 15 years of Mark-I frequency shifts are shown as well.\label{uno}}
\end{figure}
Figure~\ref{uno} shows the temporal variation of the integrated frequency 
shift for frequencies between 2 and 3.5 mHz. For comparison, the figure
includes the same calculations using the BBSO data (squares).
Mark-I instrument has been measuring the solar oscillation for more
than 20 years. This instrument observes the Sun like a star, so the 
information is limited to low degrees ($\ell \leq 3$). Recently, the
database has been re-analyzed (Jim\'enez-Reyes 2000) over a much 
wider time interval than before. 
Here, we have calculated the yearly frequency shift every six months
for the last 15 years . All the results
are very well correlated with the changes in the solar activity cycle
denoted on Fig.~\ref{uno} by the best linear fit between the radio flux 
and the results from Mark-I. 
The amplitude of these changes are, for low degrees, close to 
0.4$\mu$Hz, peak-to-peak, and the correlation with the radio flux
variations is close to 1. 

The temporal variations of the central frequency are expected to change with
both frequency and degree. Earlier works (Libbrecht and Woodard,
1990) have shown a strong variation with the frequency, the frequency shift 
being null
at $\sim$2 mHz and increasing at higher frequency. The
$\ell$-dependence has been found to be more weak.

\begin{figure}[htbp]
\centering
\includegraphics[width=.7\linewidth,angle=90]{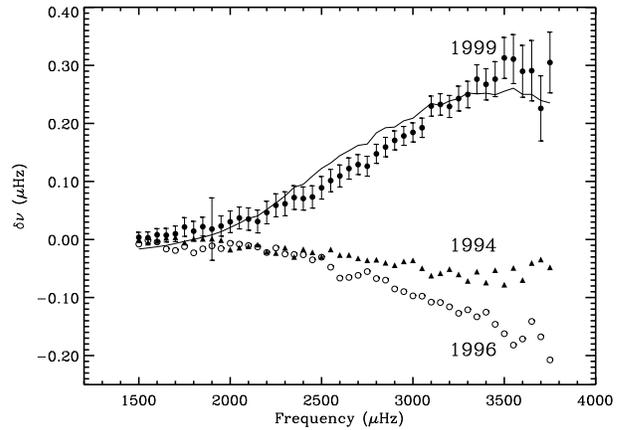}
\caption{Frequency dependence of the frequency shift for three
years. The error bars have been shown just for one year for clarity.
 The solid line represents the best linear fit
of the inverse mode mass  to the 1999 observations.
 \label{dos}}
\end{figure}

Figure~\ref{dos} shows the variation of the frequency shift averaged
over $\ell$ for three selected years. The reference is the mean over
the six years of data. The last minimum of the solar cycle happened in 1996,
whereas 1999 represents the year with higher level of solar
activity in our database. 
As we can see, the frequency dependence is very clear, mainly
between 1996 and 1999. 
\begin{figure}[htbp]
\centering
\includegraphics[width=.7\linewidth,angle=90]{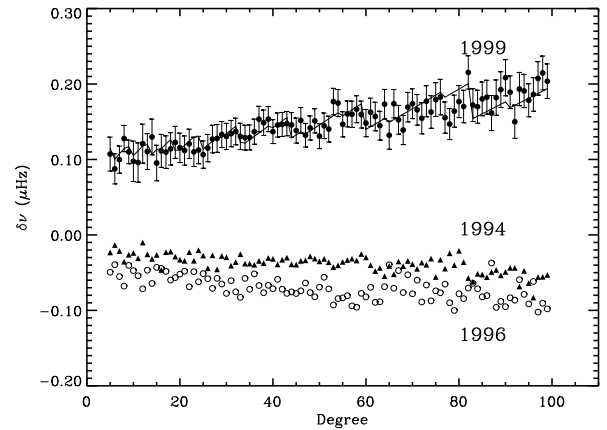}
\caption{$\ell$-dependence of the frequency shifts for three
different years. The solid line through the results corresponding
to 1999, denotes the best linear fit to the inverse mode mass.\label{tres}}
\end{figure}
The inverse mode mass has been  calculated using the model of 
Morel et al. (1997). It has been averaged in the same
way as the results shown in Figure~\ref{dos}. 
The solid line over the points
corresponding to 1999 represents the best linear fit to inverse mode mass.
The observed frequency shift exhibits almost the same frequency dependence than the inverse mode mass. However,
there are still some differences which can be seen also in the analysis
of the GONG data (Howe et al. 1999). This  may or  not be
significant but it is interesting to notice 
 that both results present the same
fluctuations around the inverse mode mass curve.

The $\ell$-dependence has been also calculated and plotted in 
Figure~\ref{tres}. Again, we have selected the same years than in the
last figure and the reference is the mean over the six years. The
dependence with the degree is maximum for 1999 with a difference 
of about $0.1\mu$Hz between $\ell$=5 and $\ell$=99.
 The inverse mode mass, averaged in the same
way than the observations, is shown in the figure, scaled to the best linear
fit using the results corresponding to 1999. The solid line follows
remarkably well the frequency shift. 

The  variations of the frequency shift
have been studied, assuming a linear relationship with the solar activity
indices. 


The solar indices used here are: 
\begin{itemize}
\item the sunspot number,R${_I}$;
\item  the integrated radio
flux at 10.7cm, F$_{10}$ \\
(both obtained from the Solar Geophysical Data);
\item  the Kitt Peak magnetic index (KPMI) extracted from the Kitt Peak full disk 
magnetograms (Harvey 1984);
\item the Mount
Wilson index also called Magnetic Plage Strength Index, MPSI 
(Ulrich et al. 1991);
\item the equivalent
width of HeI 10830\AA \ averaged over the whole solar disk using
data from Kitt Peak observatory; 
\item the Total Solar Irradiance, TSI 
(Fr\"ohlich, C. $\&$ Lean, J., 1998).
\end{itemize}

The results shown in Tab.~1 can be compared with those obtained recently by
Howe et al. (1999) and Bhatnagar et al. (1999) from GONG observations.
 The magnitude of
these parameters are very similar although some differences
remain. The reason could be the different range in frequency
taken to calculate the average. While we are taking frequencies 
between 2 and 3.5 mHz, Bhatnagar  et al. (1999)
were taking all the result between 1500 and 3500$\mu$Hz and
Howe et al. (1999) were using a mean calculated at 3mHz.

\begin{table}[htbp]
\begin{center}
\caption{Results of the  weighted linear least-squares  fits for the 
frequency shifts as a function of different solar indices.}
\begin{tabular}{lcc} 
\hline
\hline
Index & Intercept ($\mu$Hz) & Slope\\
\hline
   R$_I$  & -0.0314$\pm$0.0070 &   0.0031$\pm$0.0001$^a$ \\
  F$_{10}$  & -0.2245$\pm$0.0131 &   0.0032$\pm$0.0001$^b$\\
 KPMI  & -0.2203$\pm$0.0191 &   0.0317$\pm$0.0018$^c$\\
 MPSI  & -0.0168$\pm$0.0063 &   0.1604$\pm$0.0068$^d$\\
   He  & -0.5286$\pm$0.0416 &   0.0114$\pm$0.0008$^e$\\
TSI &-363.4905$\pm$52.6823 &   0.2662$\pm$0.0386$^f$\\
\hline
\end{tabular}
\label{table_splittings_int}
\end{center}
\end{table}
{\small
a:$\mu$Hz;\ \ \ \ \
b:$\mu$Hz/(10$^{-22}$ J/s/m$^2$/H);\ \ \ \ \ 
c,d:$\mu$Hz G$^{-1}$; \ \ \ \ \ 
e:$\mu$Hz m\AA$^{-1}$;\ \ \ \ \
f:$\mu$Hz W$^{-1}$m$^2$
}

\section{The even $a$-coefficients}

\begin{figure}[htbp]
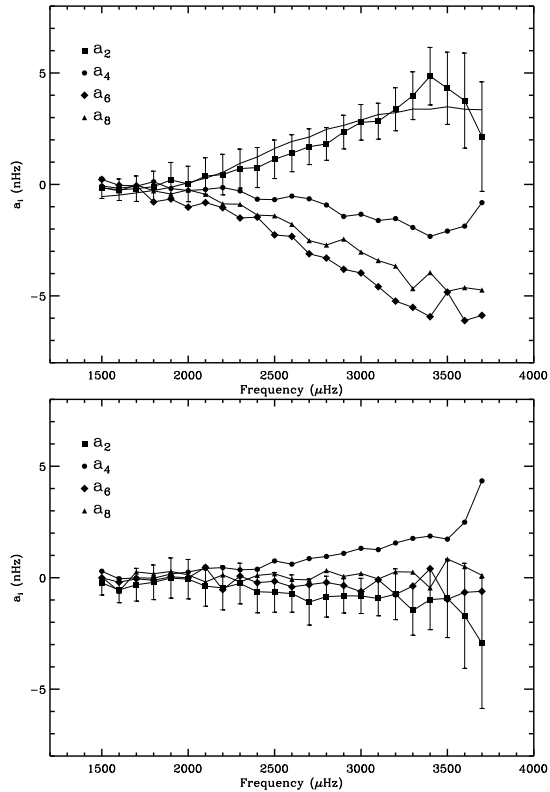

\centering
\includegraphics[width=.65\linewidth,angle=90]{af.epsi}
\includegraphics[width=.65\linewidth,angle=90]{af2.epsi}
\caption{Frequency dependence of the even $a$-coefficients for two
different years, 1996 (bottom) and 1999 (top).\label{cuatro}}
\end{figure}

The even terms are expected to change with the frequency. Therefore,
they were $\ell$-averaged over short regions in frequency, as we did
with the frequency shift. The results for two selected years, (1996 in
the bottom and 1999 in the top) are shown in Figure~\ref{cuatro}.
The even $a$-coefficients are close to zero in 1996, when the solar
cycle reached the last minimum. However, a big change can be
seen in 1999, following again the same frequency dependence than the inverse
mode mass. Again, there is a clear
fluctuation of $a_2$  around the inverse mode mass curve. Nevertheless this
may not be significant  as it stays within the error bars. 

The time-variation of the solar cycle does not affect only the central
frequency. The changes in the even $a$-coefficients demonstrate that, whatever
is the perturbation leading to these variations, it depends on the latitude.

\begin{figure*}[htbp]
\centering
\includegraphics[width=.6\linewidth,angle=90]{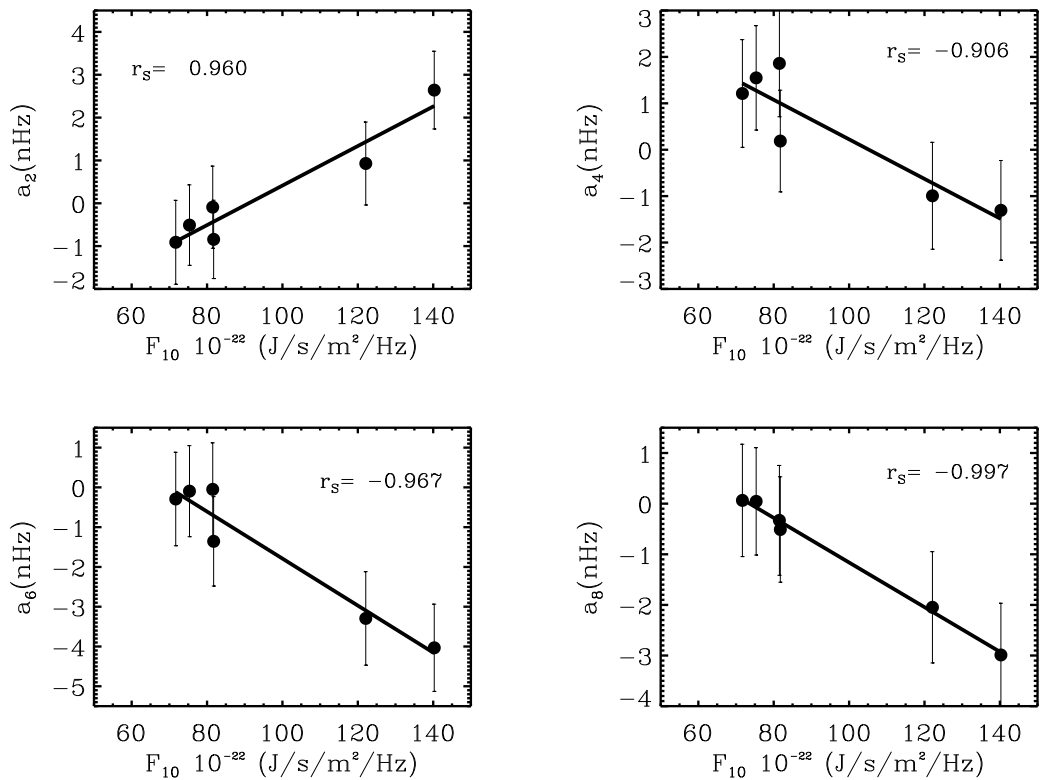}
\caption{Variation of the integrated even $a$-coefficients against the
radio flux at 10.7.cm. The solid line shows the best linear fit.\label{cinco}}
\end{figure*}

We have integrated the even terms for each one of the time series.
The resulting values are shown 
 in Figure~\ref{cinco} against the radio flux
calculated over the same period. The straight line represents the
best linear fit of both quantities. 

\section{Conclusion}
We have analyzed six years of LOWL data from 1994 to the end of 1999
in order to parameterize the variation of the frequency shift and the
even $a$-coefficients. 

In summary,
the acoustic central frequency presents a variation which is very
well correlated with the solar cycle. The integrated amplitude of this variation,
peak-to-peak, is close to 0.4$\mu$Hz. This variation is however a function
of both the frequency and the degree, and these dependences can be well fitted by the inverse
mode mass. 

In addition, the even $a$-coefficients present a significant change
with the solar cycle, very well correlated with the solar indices.
Moreover, as for the frequency shift, these changes increase with the 
frequency. 
The results agree very well with earlier studies (Libbrecht
and Woodard, 1990) carried out at the beginning of the last solar cycle.

LOWL instrument has been recently updated and a new instrument has been
developed and installed at the Observatorio del Teide (Tenerife), leading to
a new network called Experiment for Coordinated Helioseismic
Observations (ECHO). ECHO is intended to continue observing 
for a complete solar
cycle allowing us to better track the origin of the solar
activity cycle. 

\section{Acknowledgments}
We are extremely thankful to Mausumi Dikpati for useful discussions
and additional comments. S.J. Jim\'enez-Reyes and T. Corbard are very 
thankful to the organizers of the meeting for providing financial support.
T. Corbard acknowledge support from NASA grant S-92678-F.

\end{document}